\documentclass{interact}
\usepackage[utf8]{inputenc}
\usepackage{epstopdf}% To incorporate .eps illustrations using PDFLaTeX, etc.
\usepackage{subfigure}% Support for small, `sub' figures and tables
\usepackage{upgreek}%to have roman greek letters
\usepackage{xcolor}
\usepackage{soul}%to enable hihlighting with \hl

\usepackage[numbers,sort&compress,merge]{natbib}% Citation support using natbib.sty
\bibpunct[, ]{(}{)}{,}{n}{,}{,}% Citation support using natbib.sty
\begin{document}

\articletype{ARTICLE DRAFT}

\title{Absolute frequency measurements of the ${}^1{\rm S}_0 \rightarrow {}^1{\rm P}_1$ transition in ytterbium}

\author{
\name{Thomas Lauprêtre, Lucas Groult, Bachir Achi, Michael Petersen, Yann Kersalé, Marion Delehaye and Clément Lacroûte\thanks{CONTACT Cl\'ement Lacro\^ute Email: clement.lacroute@femto-st.fr}}
\affil{FEMTO-ST Institute, Université Bourgogne Franche-Comté (UBFC), CNRS, ENSMM, 26 rue de l'\' Epitaphe, 25000 Besançon, France}
}

\maketitle
% \homepage{http:...} %% author's URL, if desired

%%%%%%%%%%%%%%%%%%% abstract %%%%%%%%%%%%%%%%
%% [use \begin{abstract*}...\end{abstract*} if exempt from copyright]

\begin{abstract}
The ytterbium atom is widely used in the fields of atomic physics, cavity quantum electrodynamics, quantum information processing and optical frequency standards. There is however a strong dispersion among the reported values of the \mbox{${}^1{\rm S}_0 \rightarrow {}^1{\rm P}_1$} transition frequency. In this article, we present two independent measurements of the absolute frequency of this transition performed with two different wavemeters using atomic fluorescence spectroscopy. The cancellation of Doppler shifts is obtained by fine tuning the angle between the probe laser and the atomic beam. The resulting $^{174}\mathrm{Yb}$ isotope transition frequency is estimated to be $\,751\,526\,537\pm27$\,MHz.
\end{abstract}

%%%%%%%%%%%%%%%%%%%%%%%%%%  body  %%%%%%%%%%%%%%%%%%%%%%%%%%
\section{Introduction}

The ytterbium atom has several advantages for modern applications in atomic physics. Cold ytterbium experiments have therefore spread worldwide with applications in metrology, quantum information and quantum gases. Neutral ytterbium offers a narrow optical clock transition used for state-of-the-art frequency metrology \cite{Nemitz2016,Schioppo2017} and can be cooled down to ultra-low temperatures to study the physical properties of ultra-cold quantum gases \cite{Aguilera2018}. In addition, the Yb$^+$ ion offers two optical clock transitions at 436 nm and 467 nm \cite{Godun2014,Huntemann2016}, and its microwave clock transition at 12.6~GHz can be used for q-bit manipulation in quantum information processing experiments \cite{Randall2018}.

The  ${}^1{\rm S}_0 \rightarrow {}^1{\rm P}_1$ transition is the first cooling transition of neutral ytterbium and the first isotopic-selective stage in 2-photon ionization of ytterbium. 
However, there is a large dispersion in the measurements of the ${}^1{\rm S}_0 \rightarrow {}^1{\rm P}_1$ transition frequency~\cite{Meggers1978, Das2005, Nizamani2010,  Enomoto2016, Kleinert2016  }. The first measurement was performed in 1977 by acquiring spectrograms \cite{Meggers1978} and listed the energy of the transition as 25\,068.227\,cm$^{-1}$ (751.526\,54~THz), which was used as the reference for the NIST database~\cite{nist}. More recent measurements started in 2005, when Das \textit{et al.} used a ring-cavity resonator stabilized on a Rb transition to measure the absolute frequency in ${}^{174}{\rm Yb}$ with high accuracy \cite{Das2005}. However, the value they report, 751.525\,987\,761(60)~THz, differs from the one measured in~\cite{Meggers1978} by more than 550\,MHz. Nizamani \textit{et al.} measured the transition frequency in 2010~\cite{Nizamani2010} using a wavelength-meter specified with a 60 MHz accuracy. Their measurement, yielding the value 751.526\,65(6)\,THz, is closer to~\cite{Meggers1978} (110\,MHz difference) but not compatible with~\cite{Das2005}. In 2016, Enomoto \textit{et al.}~\cite{Enomoto2016} measured the $F=1/2\rightarrow F'=3/2$ hyperfine transition frequency in ${}^{171}{\rm Yb}$ to be 751.527\,48(10)~THz by accurately calibrating resonance frequencies of an ultra-stable Fabry-Perot cavity with lasers referenced to transitions in rubidium and calcium. They did not measure the transition frequency in the ${}^{174}{\rm Yb}$ isotope, but their result is in good agreement with~\cite{Meggers1978} and with ~\cite{Nizamani2010}, given their uncertainty and the existing isotopic shift measurements from literature~\cite{Deilamian1993,Loftus2001,Banerjee2003,Das2005,Nizamani2010,Wang2010,Kleinert2016}. Using an optical frequency comb referenced to the $\mathrm{D}_2$ line of Rb, Kleinert \textit{et al.}~\cite{Kleinert2016} made a very accurate measurement of the transition frequency in 2016, measuring $751.526\,533\,49(33)$~THz, disagreeing with the values from~\cite{Das2005} and \cite{Nizamani2010}.

Given the non-negligible dispersion of the measurements published so far, this work aims to establish and report on additional independent measurements of the absolute frequency of the ${}^1{\rm S}_0 \rightarrow {}^1{\rm P}_1$ transition in Yb. Two independent measurements of this transition for different isotopes of ytterbium are obtained by recording the fluorescence spectrum of an atomic beam. Care was taken in the experimental setup to mitigate Doppler shifts and power broadening effects, reducing systematic effects below 5~MHz. The weighted mean of our measurements results in the value $\,751\,526\,537\pm27$\,MHz for the ${}^{174}\rm{Yb}$ isotope transition frequency.%Using a rotation platform, we are able to cancel the Doppler shift to less than 5~MHz.

%%%%%%%%%%%%%%%%%%%%%%%%%%%%%%%%%%%%%%%%%%%%%%%%%%%%%%%%%%%%%%%%%%%%%%%%
\section{Experimental setup}\label{setup}

\begin{figure}[h!]
	\centering
	\includegraphics[width=0.3\linewidth]{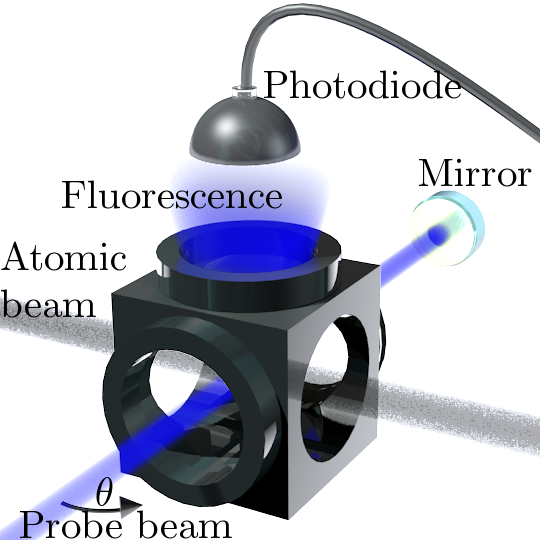}
	\caption{Sketch of the experimental setup. In an ultra-high vacuum chamber, a 399 nm laser beam crosses an atomic beam of neutral ytterbium atoms to excite the ${}^1{\rm S}_0 \rightarrow {}^1{\rm P}_1$ transition. The resulting fluorescence is detected by a low-noise photodiode. An angle $\theta$ can be introduced with respect to the perfect orthogonal configuration and a mirror allows for back reflection of the laser beam for alignment purposes (see text for details).}
	\label{fig:ybneutral-setupv3}
\end{figure}

The experimental set-up is sketched out in Fig.~\ref{fig:ybneutral-setupv3}. A commercial dispenser of ytterbium with natural isotopic abundance\cite{Alvasource} is driven at 5.5\,A and thus heated close to 600$^{\circ}$C in order to produce an atomic beam inside an ultra-high vacuum chamber (background pressure $< 10^{-10}$~mbars). The  ${}^1{\rm S}_0 \rightarrow {}^1{\rm P}_1$ transition in ytterbium is excited by a 399\,nm laser beam crossing the atomic beam nearly orthogonally. The laser light, produced by a commercial external-cavity diode laser (ECDL)\cite{Toptica} with a linewidth of a few MHz, is brought to the experiment chamber via a single-mode fiber. The fiber output coupler is mounted on a rotatable breadboard in order to adjust the angle $\theta$ between the probe laser and the direction perpendicular to the atomic beam. The average intensity of the probe beam is $\sim 8~\mathrm{mW.cm^{-2}}$ corresponding to a saturation parameter of 0.14. The laser frequency is slowly scanned over a $\sim 3~\mathrm{GHz}$ range using the piezoelectric actuator driving its grating (PZT) during approximately $100~\mathrm{s}$. High-apperture optics collect photons scattered by atoms excited to the upper state and decaying back to the ground state. The fluorescence signal is detected with a photodiode\cite{Thorlabs}. While the fluorescence spectrum of the transition is recorded, the frequency of the laser is sampled every $\sim 25~\mathrm{ms}$ on two independent commercial wavemeters\cite{HighFinesse}. The first wavemeter is a WS/7 Super Precision specified by the manufacturer with an absolute accuracy of $\pm 60~\mathrm{MHz}$ when calibrated with its built-in neon lamp. Its frequency drift in absence of thermal isolation has been measured to be below 30\,MHz/day \cite{Saleh2015}. The second wavemeter is a WS8-2 calibrated to the $F=3\rightarrow F'=4$ hyperfine transition of the $\mathrm{D}_1$ line of caesium at 894\,nm, which results in an absolute accuracy of $\pm30$\,MHz at 399\,nm as specified by the manufacturer. 

A mirror placed at the output of the chamber can be used to reflect the laser beam back onto itself. In that case, the two counter-propagating beams have perfectly opposite Doppler shifts with respect to the atomic beam for non-zero angles $\theta$, allowing to tune $\theta$ to zero with mrad precision (see section~\ref{uncertainty}).

%%%%%%%%%%%%%%%%%%%%%%%%%%%%%%%%%%%%%%%%%%%%%%%%%%%%%%%%%%%%%%%%%%%%%%%%%%
\section{Results}

\begin{figure}[h!]
	\centering
	\includegraphics[width=0.5\linewidth]{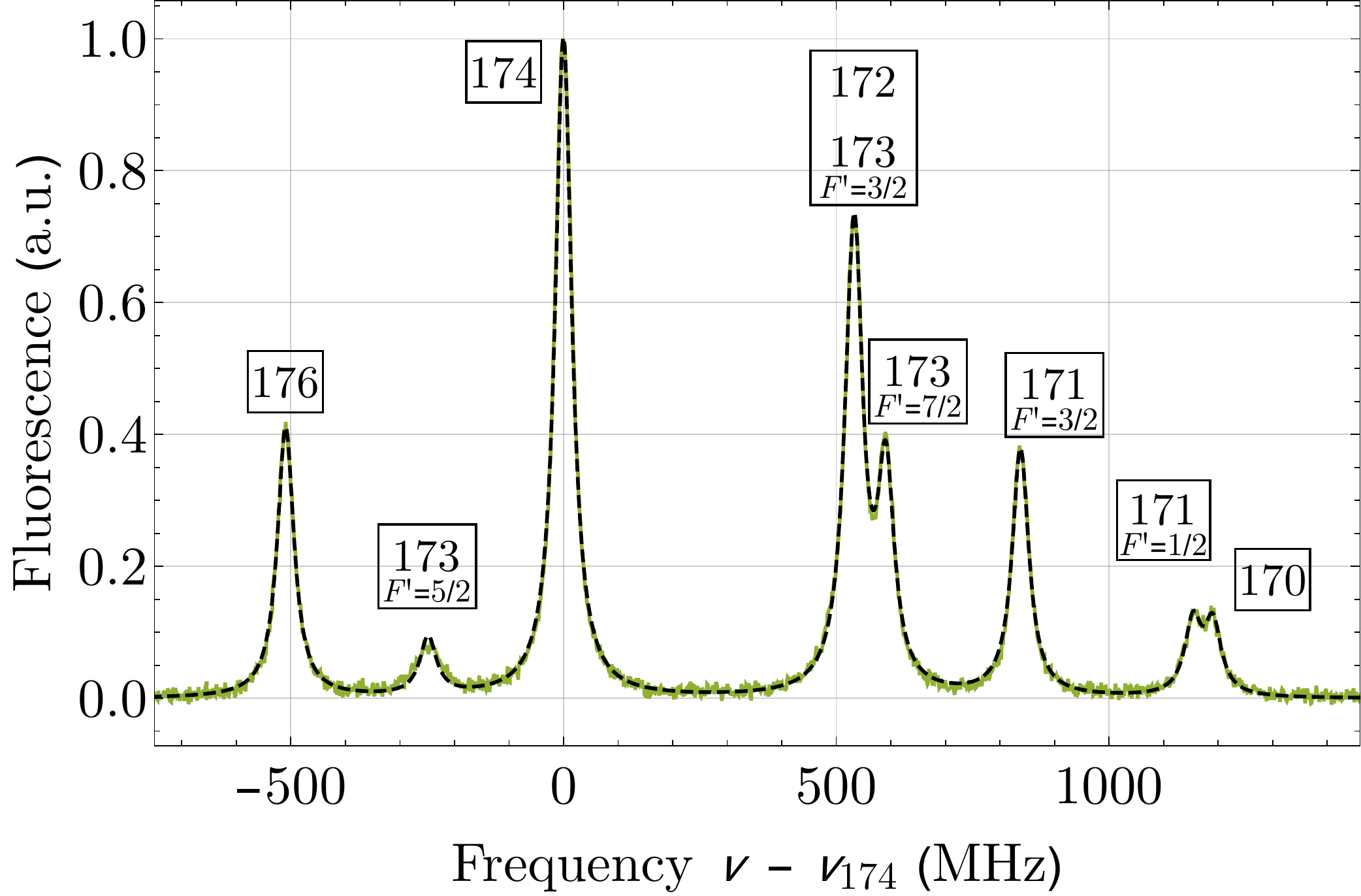}
	\caption{Fluorescence spectrum of ytterbium atoms, where the laser frequency is measured with the WS8-2. Green points: experimental data. Black dashed curve: Lorentzian fit with common FWHM. Frequencies are plotted relatively to the ${}^{174}\rm{Yb}$ isotope line.}
	\label{fig:spectrum2}
\end{figure}

The fluorescence spectrum was recorded at $\theta=0^{\circ}$ as a function of the frequency for each wavemeter. %, by linearly interpolating the frequency measurements as a function of time and using a common time reference provided by unblocking the laser light. 
A Lorentzian fit is applied to the data to extract the frequency of the ${}^1{\rm S}_0 \rightarrow {}^1{\rm P}_1$ transition for each of the ytterbium isotopes. Fig.~\ref{fig:spectrum2} shows the spectrum obtained with the WS8-2. The common estimated full width at half-maximum (FWHM) of 37~MHz is larger than the 28~MHz natural linewidth. This minor broadening is mainly attributed to laser linewidth ($\sim 4~\mathrm{MHz}$), power broadening ($\sim 2~\mathrm{MHz}$) and Doppler broadening due to the transverse velocity distribution in the atomic beam.

%%%%%%%%%%%%%%%%%%%%%%%%%%%%%%%%%%%%%%
\subsection{Transition frequency of the ${}^{174}\rm{Yb}$ isotope}

\begin{figure}[h!]
	\centering
	\includegraphics[width=0.7\linewidth]{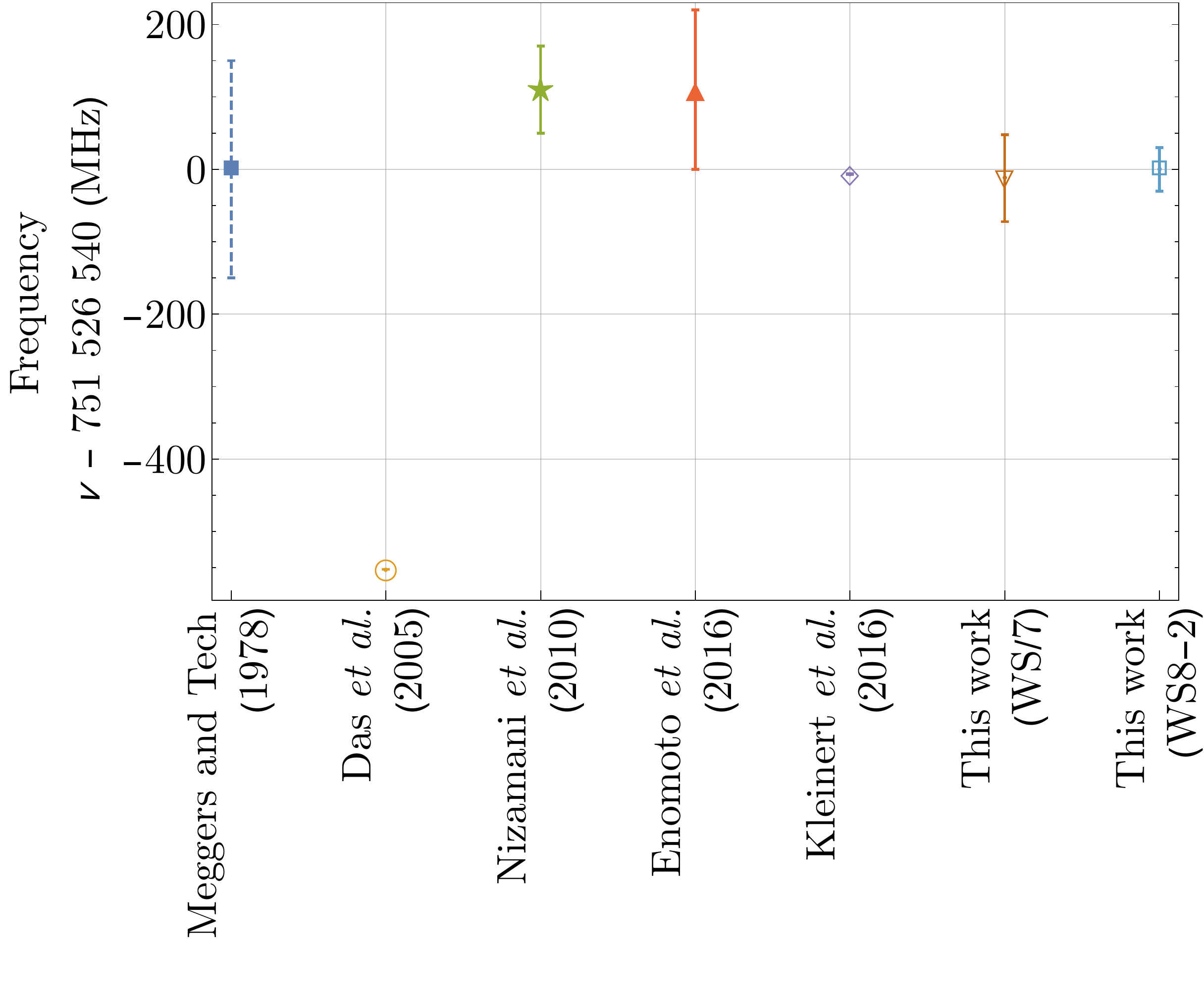}
	\caption{Absolute measurements of the ${}^{174}\rm{Yb}$ ${}^1{\rm S}_0 \rightarrow {}^1{\rm P}_1$ transition frequency. Blue square:~\cite{Meggers1978}; yellow circle:~\cite{Das2005}; green star:~\cite{Nizamani2010}; red upward pointing triangle: calculated from~\cite{Enomoto2016}; purple diamond:~\cite{Kleinert2016}; brown downward pointing triangle: this work with WS/7; light blue square: this work with WS8-2. Frequencies are plotted relatively to the atomic transition in the ${}^{174}\rm{Yb}$ isotope as measured in \cite{Meggers1978}.}
	\label{fig:comparison}
\end{figure}

Our measurement, including all sources of uncertainty (see discussion section~\ref{uncertainty}), returns the value $751\,526\,528\pm60$\,MHz for the frequency of the ${}^1{\rm S}_0 \rightarrow {}^1{\rm P}_1$ transition in the ${}^{174}\rm{Yb}$ isotope when measured with the WS/7, and $751\,526\,540\pm30$\,MHz when measured with the WS8-2. These results are compared with previously reported values in Fig.~\ref{fig:comparison}. No uncertainty is provided for the measurement from~\cite{Meggers1978}, but Kleinert \textit{et al.} gave a $\pm150$\,MHz estimate for this datapoint from discussions with the NIST team \cite{Kleinert2016}. In~\cite{Enomoto2016}, the authors do not provide a frequency value for the ${}^{174}{\rm Yb}$ isotope. For comparison purposes, we use the value of $751\,526.65\pm0.11$\,GHz calculated from their measurement for the ${}^{171}{\rm Yb}$ isotope translated by an isotopic shift of 832.6\,MHz, the weighted mean \cite{Weights} of the shifts measurements from~\cite{Deilamian1993,Loftus2001,Banerjee2003,Das2005,Nizamani2010,Wang2010,Kleinert2016}. An uncertainty of 10\,MHz was added to account for the uncertainty on the isotopic shifts measurements.

The reported values are spread out over 600 MHz. When restraining the analysis to values from \cite{Meggers1978,Nizamani2010,Enomoto2016,Kleinert2016} as well as ours, we find a reasonable agreement, with error bars still covering nearly 400 MHz. Within the error bars, our results are compatible with \cite{Meggers1978,Enomoto2016,Kleinert2016}, and our WS8-2 measurement differs from \cite{Nizamani2010} by a few MHz only. By doing a simple averaging of all 6 values, we find $\bar{\nu}_{174}=\,751\,526\,574\pm35$\,MHz, while a weighted mean\cite{Weights} yields $\tilde{\nu}_{174}=\,751\,526\,533.5\pm0.3$\,MHz.

%%%%%%%%%%%%%%%%%%%%%%%%%%%%%%%%%%%%%%%%%
\subsection{Other isotopes}

Table~\ref{tbl:isotopes} displays our measurements of the ytterbium ${}^1{\rm S}_0 \rightarrow {}^1{\rm P}_1$ transition absolute frequencies for the different isotopes and hyperfine transitions. The uncertainty of the wavemeters when measuring a frequency difference over the $\sim$GHz range is specified by the manufacturer to be equal to the absolute accuracy. For this reason, we do not find necessary to present a thorough comparison of our results with the isotopic shifts measurements found in literature, which have a much smaller dispersion than this uncertainty. Our results are nevertheless in very good agreement with all the isotopic shifts measurements found in ~\cite{Deilamian1993,Loftus2001,Banerjee2003,Das2005,Nizamani2010,Wang2010,Kleinert2016}.

It should be noted that even though the wavemeters manufacturer recommends using the absolute uncertainty for frequency difference measurements, this seems to be a too conservative estimate. To verify this, we successively locked an 870~nm ECDL to adjacent teeth of a frequency comb. The teeth had a fixed spacing of about 250~MHz and stable well below the MHz level. By reading the frequency-doubled output of this laser with the wavemeters after each 250 MHz step, we checked that the uncertainty on frequency difference measurements was below 6~MHz for both the WS/7 and WS8-2 wavemeters across  a $\sim$7~GHz range around 435~nm. We are unfortunately unable to perform this test around 399 nm to verify that our measurement could be used to estimate the isotope shifts at the few MHz level.

\begin{table}[h!]
\centering
	\begin{tabular}{| c | c | c | c |}
				\hline
		& \multicolumn{3}{c|}{Transition frequency} \\
		\cline{2-4}
		Isotope 	&  WS/7 (MHz) &  WS8-2 (MHz) & Weighted Mean (MHz)\\ \hline\hline

		${}^{176}\mathrm{Yb}$		&  751\,526\,021(60) &	751\,526\,031(30)	&	751\,526\,029(27)	\\ \hline

		${}^{173}\mathrm{Yb}	$	& 751\,526\,280(60) &	751\,526\,292(30) &	751\,526\,289(27) \\ \hline
		
		${}^{172}\mathrm{Yb}$ \&
		${}^{173}\mathrm{Yb}$
	$\scriptstyle{({\rm F'}=3/2)}$	& 751\,527\,063(60) &	751\,527\,072(30) &	751\,527\,070(27)	\\ \hline		

		${}^{173}\mathrm{Yb}$
	$\scriptstyle{({\rm F'}=7/2)}$	& 751\,527\,122(60) &	751\,527\,131(30) &	751\,527\,129(27)		 \\ \hline
		
		${}^{171}\mathrm{Yb}$
	$\scriptstyle{({\rm F'}=3/2)}$	& 751\,527\,370(60)	&	751\,527\,378(30) &	751\,527\,376(27)	 \\ \hline

		${}^{171}\mathrm{Yb}$
	$\scriptstyle{({\rm F'}=1/2)}$	& 751\,527\,689(60) &	751\,527\,694(30) &	751\,527\,693(27)		\\ \hline
		
		${}^{170}\mathrm{Yb}$		& 751\,527\,725(60) &	751\,527\,730(30)  &	751\,527\,729(27)	\\
		 \hline
	\end{tabular}
\caption{Absolute measurement of the frequency of the ${}^1{\rm S}_0 \rightarrow {}^1{\rm P}_1$ transition depending on the isotope and the hyperfine transition. A single measurement is provided for ${}^{172}\mathrm{Yb}$ and $F=5/2\rightarrow F'=3/2$ in ${}^{173}\mathrm{Yb}$ as the two resonances are not distinguishable.}
\label{tbl:isotopes}
\end{table}

%%%%%%%%%%%%%%%%%%%%%%%%%%%%%%%%%%%%%%%%%%%%%%%%%%%%%%%%%%%%%%%%%%
\section{Sources of uncertainty}\label{uncertainty}

\subsection{Measurement method and frequency scale}

We record fluorescence spectra by acquiring a time trace of the fluorescence signal with an oscilloscope while simultaneously monitoring the laser frequency with the wavemeters. During a measurement the 100~s-long frequency scan is made over a range of 3\,GHz, which induces uncalibrated non-linearity when using the laser PZT. In practice, we sample the laser frequency every $\sim 25~\mathrm{ms}$ using the wavemeters and interpolate between the measurements to convert the oscilloscope time trace to a frequency scale. The error on the frequency scale origin from this conversion results in an uncertainty corresponding to one time step at maximum, and is thus evaluated to be below 0.8~MHz.

\subsection{Systematic errors}

In addition to the uncertainty inherent to the measurement method, different sources of error of physical origin can be identified, including frequency shifts and broadenings due to the laser intensity and linewidth, the residual magnetic fields and the laser polarization, and the Doppler effect.

First, the resonances in the recorded spectrum have no detectable amount of asymmetry. The position of their maxima stays within the error bar returned by the fit, which is found to be sub-MHz for all of them. It is therefore very conservative to assume an error of $\pm$1\,MHz from the combined effects leading to a broadening of the resonances, including non-linearity of the frequency measurement over the linewidth range.

The experiment, not protected by a magnetic shield, is subject to the earth field and environmental background fields. The residual magnetic field around the chamber was measured to have an amplitude below 1\,G leading to a splitting  of the transitions of less than 1.4\,MHz for both odd and even isotopes. Moreover, the laser polarization is random. In these conditions, Zeeman splitting leads to a minor asymmetric broadening of the recorded resonances (natural linewidth of 28 MHz) and sub-MHz shifts of the maxima. For odd isotopes, quantum interferences and polarization effects~\cite{Kleinert2016,Brown2013} can also lead to sub-MHz shifts of the resonances. In the particular case of the $F=5/2\rightarrow F'=3/2$ hyperfine transition in ${}^{173}$Yb, very specific polarizations of the laser with respect to the residual magnetic field would result in an optical pumping effect \cite{McKnight2018} and forbid detection of fluorescence for this transition. We do not claim in this work to measure separately transition frequencies for ${}^{172}$Yb and ${}^{173}$Yb${}_{F=5/2\rightarrow F'=3/2}$, and the presence or the absence of the ${}^{173}$Yb${}_{F=5/2\rightarrow F'=3/2}$ peak in the fluorescence spectrum can only induce sub-MHz level uncertainty on the common measurement largely dominated by the ${}^{172}$Yb transition frequency value. In conclusion, residual magnetic fields and polarization effects are conservatively considered to induce an overall error of $\pm 3~\mathrm{MHz}$ on the transition frequency.

Doppler shifts can be caused by the non-perpendicularity of the laser beam relatively to the atomic beam direction, which is not perfectly known inside the vacuum chamber. The longitudinal velocity distribution of the atomic beam is centered around a non-zero value, which leads to a Doppler shift depending on the angle $\theta$. From the temperature of the dispenser $T\sim875$\,K, we can roughly estimate this shift\cite{Speed} to $\frac{vel_{\rm{at}}\sin{\theta}}{\lambda}\sim\sin{\theta}\times725\,\mathrm{MHz}$, with $vel_{\rm{at}}$ the most probable speed and $\lambda$ the laser wavelength. This amounts to $\sim13\,\mathrm{MHz/}^{\circ}$ at small angles. This sensitivity can thus induce a significant contribution to the systematic error budget. In order to reduce the uncertainty due to this effect, we recorded the fluorescence spectrum at different angles $\theta$ using the previously described retro-reflection mirror (section~\ref{setup} and Fig.~\ref{fig:ybneutral-setupv3}).

\begin{figure}[h!]
	\centering

	\includegraphics[width=0.5\linewidth]{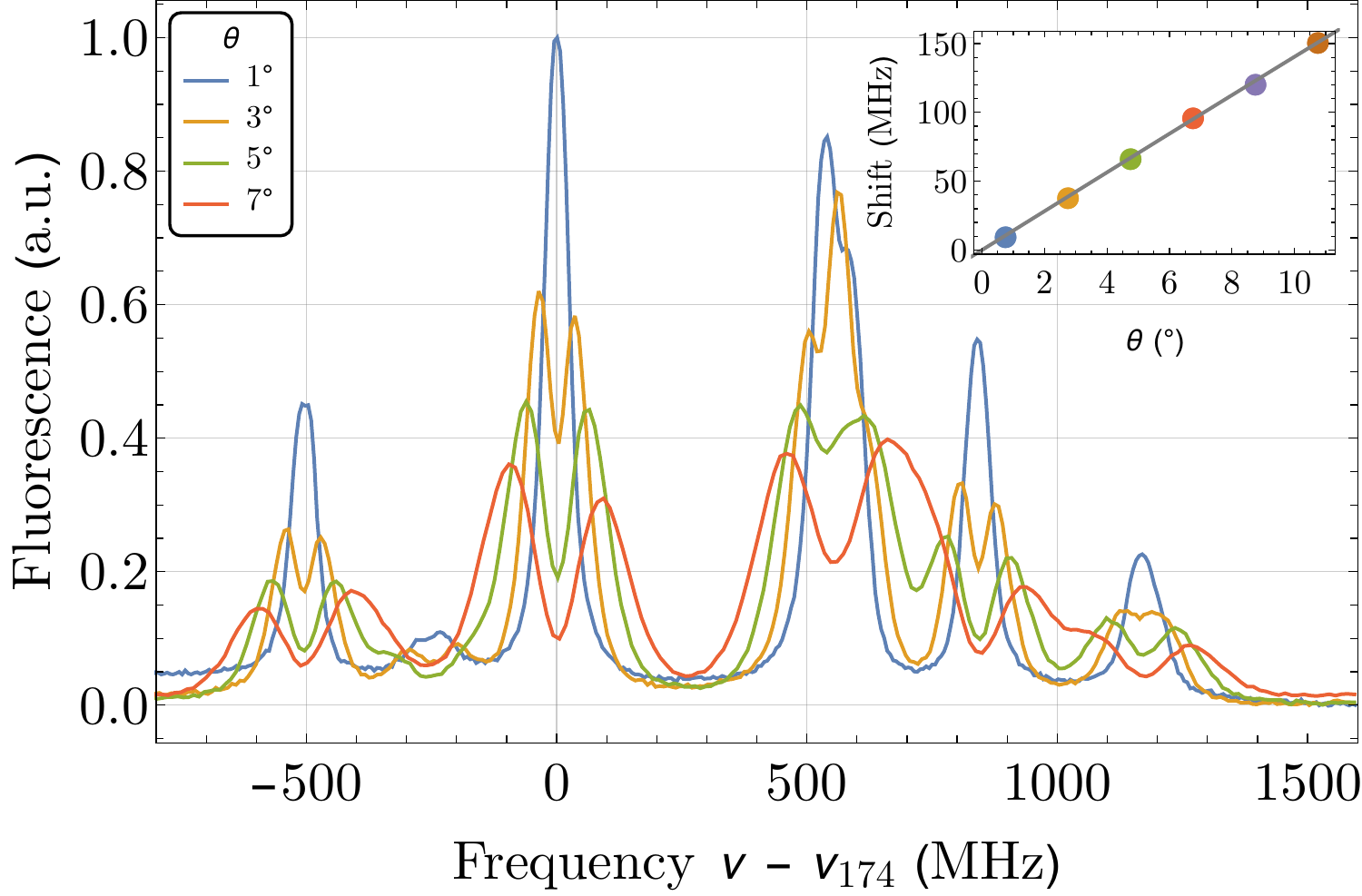}
	\caption{Fluorescence spectra of ytterbium atoms as a function of the laser frequency while retro-reflecting the probing beam at different angles: $\theta\sim1^{\circ}$ (blue), $\theta\sim3^{\circ}$ (yellow), $\theta\sim5^{\circ}$ (green), $\theta\sim7^{\circ}$ (red). Frequencies are plotted relatively to the atomic transition in the ${}^{174}\rm{Yb}$ isotope. Inset: Doppler shift as a function of the angle $\theta$ and associated linear fit indicating the zero-shift angle for perpendicular configuration.}
	\label{fig:all}
\end{figure}

Figure~\ref{fig:all} shows recorded fluorescence spectra for different values of the angle $\theta$. For each $\theta$ value, each resonance peak is split into a doublet of separate resonances corresponding to opposite Doppler shifts. For clarity, spectra have been  normalized to the spectrum obtained with $\theta=1^{\circ}$. A horizontal average has also been applied in order to smooth out the curves for better readability. For analysis, the data have been fitted with the sum of two translated spectra which include all isotopes lines (one for each way of the laser propagation). In the fitting function, each line is considered a Voigt profile, which reduces processing to an acceptable time, and the symmetrical Doppler shift of each of the two spectra is left as a common free parameter. Error bars returned by the fits are all sub-MHz regarding the Doppler shift results. These shifts are then plotted as a function of the angle $\theta$ and finally linearly fitted as a function of $\sin{\theta}$ in order to find the position for which the laser beam is perpendicular to the most probable velocity, \emph{ie} $\theta=0^{\circ}$. These results are presented in the inset of Fig.~\ref{fig:all}. The 1${}^{\circ}$ data point has been discarded in this last analysis even though the fitting process of the recorded spectrum seems to return a reliable result. The fit indicates a $\sin{\theta}\times(787\pm5)\,\mathrm{MHz}$ dependence of the Doppler shift. Given the error on the perpendicular position from the analysis ($\pm 0.3^{\circ}$) and the angular resolution of the rotatable breadboard (5'$\sim 0.083{}^{\circ}$), we estimate the uncertainty on frequency measurements caused by this parameter to be $\pm 4.3$\,MHz. Note that an asymmetrical Doppler broadening due to the same effect is to be expected on the final recorded spectrum, but this has also been minimized by the above procedure.

\subsection{Overall uncertainty}

All pre-mentionned uncertainties have been summarized in Table~\ref{tbl:uncertainty_budget} as well as the accuracy for each wavemeter. From these results, we deduce an overall uncertainty of $\pm$60\,MHz for the WS/7 measurement and $\pm$30\,MHz for the WS8-2. With all other sources of uncertainties having a very small contribution, the overall uncertainty is completely determined by the wavemeters. 

Using these uncertainties and deriving the weighted mean of our two results, the absolute frequency of the ${}^1{\rm S}_0 \rightarrow {}^1{\rm P}_1$ transition of ${}^{174}\rm{Yb}$ is found to be $\,751\,526\,537\pm27$\,MHz. For other isotopes, the weighted averages have been indicated in Table~\ref{tbl:isotopes}.

\begin{table}[h!]
\centering
	\begin{tabular}{| l | c |}
			\hline
		Source of error 	&  Uncertainty ($\pm$MHz) \\ \hline\hline

		Frequency scale conversion													&  0.8 	\\ \hline

		Line broadening effects															& 1  \\ \hline
		
		Residual magnetic field	and laser polarization			& 3 	\\ \hline	
		
		Laser/atomic beams non orthogonality								& 4.3 	\\ \hline
				
		Wavemeter WS/7																			& 60 	\\ \hline
				
		Wavemeter WS8-2																			& 30 	\\ \hline

		 \hline
	\end{tabular}
\caption{Summary of the conservative estimates on the uncertainties affecting the ${}^1{\rm S}_0 \rightarrow {}^1{\rm P}_1$ transition measurement.}
\label{tbl:uncertainty_budget}
\end{table}

%%%%%%%%%%%%%%%%%%%%%%%%%%%%%%%%%%%%%%%%%%%%%%%%%%%%%%%%%%%%%%%%%%%%%%%%
\section{Conclusion}

In this article, we have reported on two independent measurements of the ${}^1{\rm S}_0 \rightarrow {}^1{\rm P}_1$ transition frequency for different isotopes of neutral ytterbium. The measurements were done by recording fluorescence spectra using two independent commercial wavemeters. We implemented a method based on counterpropagating beams that minimizes the uncertainty due to the Doppler shift, by identifying the angle for which the laser beam is perpendicular to the atomic beam. 

By presenting these results, we hope to reduce the existing uncertainty on this transition frequency, suffering from a non-negligible dispersion in the literature. Our results confirm in particular the most recent accurate measurement by Kleinert \textit{et al.}~\cite{Kleinert2016} from 2016. This might be useful to future starting experiments based on neutral and ionized ytterbium where knowing accurately this transition frequency is crucial.

\section*{Funding}

This work has been supported by the Agence Nationale de la Recherche (ANR-14-CE26-0031-01 MITICC, ANR-10-LABX-48-01 First-TF, and ANR-11-EQPX-0033 Oscillator-IMP), the Région Bourgogne Franche-Comté, the Centre National d'\' Etudes Spatiales and the EIPHI Graduate School (contract "ANR-17-EURE-0002").

\section*{Acknowledgments}
The authors would like to thank Emmanuel Bigler for designing and setting up the detection optics, Vincent Giordano for fruitful discussions about the experiment, Jacques Millo for constructive remarks and help with the optical frequency comb operation, and Rodolphe Boudot for his precious comments about the wavemeter calibration and the careful reading of the manuscript.

\section*{Disclosures}
The authors declare that there are no conflicts of interest related to this article.

%\renewcommand\refname{References and Notes}
%%%%%%%%%% If using BibTeX:
\bibliographystyle{unsrt}
\bibliography{Biblio}

\begin{thebibliography}{10}

\bibitem{Nemitz2016}
Nils Nemitz, Takuya Ohkubo, Masao Takamoto, Ichiro Ushijima, Manoj Das, Noriaki
  Ohmae, and Hidetoshi Katori.
\newblock Frequency ratio of {Y}b and {S}r clocks with 5 $\times$ 10${}^{-17}$
  uncertainty at 150 seconds averaging time.
\newblock {\em Nature Photonics}, 10(4):258–261, Feb 2016.

\bibitem{Schioppo2017}
M.~Schioppo, R.~C. Brown, W.~F. McGrew, N.~Hinkley, R.~J. Fasano, K.~Beloy,
  T.~H. Yoon, G.~Milani, D.~Nicolodi, J.~A. Sherman, N.~B. Phillips, C.~W.
  Oates, and A.~D. Ludlow.
\newblock Ultrastable optical clock with two cold-atom ensembles.
\newblock {\em Nature Photonics}, 11(1):48--52, January 2017.

\bibitem{Aguilera2018}
M.~Bosch Aguilera, R.~Bouganne, A.~Dareau, M.~Scholl, Q.~Beaufils, J.~Beugnon,
  and F.~Gerbier.
\newblock Non-linear relaxation of interacting bosons coherently driven on a
  narrow optical transition.
\newblock {\em EPL (Europhysics Letters)}, 123(4):40004, September 2018.

\bibitem{Godun2014}
R.~M. Godun, P.~B.~R. Nisbet-Jones, J.~M. Jones, S.~A. King, L.~A.~M. Johnson,
  H.~S. Margolis, K.~Szymaniec, S.~N. Lea, K.~Bongs, and P.~Gill.
\newblock Frequency ratio of two optical clock transitions in
  ${}^{171}${Y}b${}^{+}$ and constraints on the time variation of fundamental
  constants.
\newblock {\em Physical Review Letters}, 113(21):210801, November 2014.

\bibitem{Huntemann2016}
N.~Huntemann, C.~Sanner, B.~Lipphardt, Chr. Tamm, and E.~Peik.
\newblock Single-ion atomic clock with
  $3\ifmmode\times\else\texttimes\fi{}{10}^{-18}$ systematic uncertainty.
\newblock {\em Physical Review Letters}, 116(6):063001, February 2016.

\bibitem{Randall2018}
J.~Randall, A.~M. Lawrence, S.~C. Webster, S.~Weidt, N.~V. Vitanov, and W.~K.
  Hensinger.
\newblock Generation of high-fidelity quantum control methods for multilevel
  systems.
\newblock {\em Physical Review A}, 98(4):043414, October 2018.

\bibitem{Meggers1978}
W.~F. Meggers and J.~L. Tech.
\newblock The first spectrum of ytterbium ({Yb I}).
\newblock {\em J. Res. Bur. Stand.}, 83(1):13, 1978.

\bibitem{Das2005}
Dipankar Das, Sachin Barthwal, Ayan Banerjee, and Vasant Natarajan.
\newblock Absolute frequency measurements in {Yb} with
  $0.08\phantom{\rule{0.3em}{0ex}}\mathrm{ppb}$ uncertainty: Isotope shifts and
  hyperfine structure in the $399\text{\ensuremath{-}}\mathrm{nm}$
  $^{1}${S}$_{0}\ensuremath{\rightarrow}^{1}${P}$_{1}$ line.
\newblock {\em Phys. Rev. A}, 72:032506, Sep 2005.

\bibitem{Nizamani2010}
Altaf~H. Nizamani, James~J. McLoughlin, and Winfried~K. Hensinger.
\newblock Doppler-free {Y}b spectroscopy with the fluorescence spot technique.
\newblock {\em Phys. Rev. A}, 82:043408, Oct 2010.

\bibitem{Enomoto2016}
Katsunari Enomoto, Nagisa Hizawa, Takahiro Suzuki, Kaori Kobayashi, and Yoshiki
  Moriwaki.
\newblock Comparison of resonance frequencies of major atomic lines in 398--423
  nm.
\newblock {\em Applied Physics B}, 122(5):126, Apr 2016.

\bibitem{nist}
A.~Kramida, Yu. Ralchenko, J.~Reader, and NIST~ASD Team.
\newblock {NIST Atomic Spectra Database (Ver. 5.6) (National Institute of
  Standards and Technology, Gaithersburg, 2018). http://physics.nist.gov/asd}.

\bibitem{Deilamian1993}
K.~Deilamian, J.~D. Gillaspy, and D.~E. Kelleher.
\newblock Isotope shifts and hyperfine splittings of the 398.8-nm {Yb I} line.
\newblock {\em J. Opt. Soc. Am. B}, 10(5):789--793, May 1993.

\bibitem{Loftus2001}
T.~Loftus, J.~R. Bochinski, and T.~W. Mossberg.
\newblock Optical double-resonance cooled-atom spectroscopy.
\newblock {\em Phys. Rev. A}, 63:023402, Jan 2001.

\bibitem{Banerjee2003}
A~Banerjee, U.~D Rapol, D~Das, A~Krishna, and V~Natarajan.
\newblock Precise measurements of {UV} atomic lines: Hyperfine structure and
  isotope shifts in the 398.8 nm line of {Y}b.
\newblock {\em Europhysics Letters ({EPL})}, 63(3):340--346, aug 2003.

\bibitem{Wang2010}
Wen-Li Wang and Xin-Ye Xu.
\newblock A novel method to measure the isotope shifts and hyperfine splittings
  of all ytterbium isotopes for a 399-nm transition.
\newblock {\em Chinese Physics B}, 19(12):123202, dec 2010.

\bibitem{Kleinert2016}
Michaela Kleinert, M.~E. Gold~Dahl, and Scott Bergeson.
\newblock Measurement of the {Yb I}
  $^{1}${S}$_{0}\text{\ensuremath{-}}^{1}${P}$_{1}$ transition frequency at 399
  nm using an optical frequency comb.
\newblock {\em Phys. Rev. A}, 94:052511, Nov 2016.

\bibitem{Alvasource}
Alvasources from Alvatec (now AlfaVakuo e.U.).

\bibitem{Toptica}
Toptica DL100.

\bibitem{Thorlabs}
Thorlabs PDF10A/M.

\bibitem{HighFinesse}
HighFinesse WS/7 and WS8-2.

\bibitem{Saleh2015}
Khaldoun Saleh, Jacques Millo, Alexandre Didier, Yann Kersal{\'{e}}, and
  Cl{\'{e}}ment Lacro{\^{u}}te.
\newblock Frequency stability of a wavelength meter and applications to laser
  frequency stabilization.
\newblock {\em Applied Optics}, 54(32):9446, nov 2015.

\bibitem{Weights}
Using the inverse of the squared uncertainties as weights.

\bibitem{Brown2013}
Roger~C. Brown, Saijun Wu, J.~V. Porto, Craig~J. Sansonetti, C.~E. Simien,
  Samuel~M. Brewer, Joseph~N. Tan, and J.~D. Gillaspy.
\newblock {Quantum interference and light polarization effects in unresolvable
  atomic lines: Application to a precise measurement of the ${}^{6,7}$Li
  ${D}_{2}$ lines}.
\newblock {\em Phys. Rev. A}, 87:032504, Mar 2013.

\bibitem{McKnight2018}
Quinton McKnight, Adam Dodson, Tucker Sprenkle, Tyler Bennett, and Scott
  Bergeson.
\newblock Comment on ``laser cooling of $^{173}\mathrm{Yb}$ for isotope
  separation and precision hyperfine spectroscopy''.
\newblock {\em Phys. Rev. A}, 97:016501, Jan 2018.

\bibitem{Speed}
Considering a Maxwell-Boltzmann distribution with most probable speed
  $vel_{\rm{at}}=\sqrt{\frac{2k_\mathrm{B}T}{M}}$, with $k_\mathrm{B}$ the
  Boltzmann constant and $M$ the mass of an ytterbium atom.

\end{thebibliography}

\end{document}